\documentclass[[final,3p,times,twocolumn]{elsarticle}
\usepackage{graphicx}
\usepackage{dcolumn}
\usepackage{bm}
\usepackage{amsmath}
\usepackage{comment}
\usepackage{epstopdf} 
\usepackage{xr-hyper}
\usepackage{lineno,hyperref}
\usepackage{color} 
\usepackage{soul} 

\makeatletter
\newcommand*{\addFileDependency}[1]{
	\typeout{(#1)}
	\@addtofilelist{#1}
	\IfFileExists{#1}{}{\typeout{No file #1.}}
}
\makeatother

\usepackage{amssymb}

\usepackage{lineno,hyperref}
\modulolinenumbers[5]

\journal{Ultramicroscopy}

\begin{document}

\begin{frontmatter}



\title{Reducing electron beam damage through alternative STEM scanning strategies. Part II -- Attempt towards an empirical model describing the damage process}

\author[emat,nanolab]{D. Jannis\fnref{equal}}
\author[emat,nanolab]{A. Velazco\fnref{equal}}
\fntext[equal]{Both first authors contributed equally.}
\author[emat,nanolab]{A. B\'ech\'e}
\author[emat,nanolab]{J. Verbeeck\corref{mycorrespondingauthor}}
\cortext[mycorrespondingauthor]{Corresponding author}
\ead{jo.verbeeck@uantwerp.be}

\address[emat]{EMAT, University of Antwerp, Groenenborgerlaan 171, 2020 Antwerp, Belgium}
\address[nanolab]{NANOlab Center of Excellence, University of Antwerp, Groenenborgerlaan 171, 2020 Antwerp, Belgium}

\begin{abstract}
In this second part of a series we attempt to construct an empirical model that can mimick all experimental observations made regarding the role of an alternative interleaved scan pattern in STEM imaging on the beam damage in a specific zeolite sample. We make use of a 2D diffusion model that describes the dissipation of the deposited beam energy in the sequence of probe positions that are visited during the scan pattern. The diffusion process allows for the concept of trying to 'outrun' the beam damage by carefully tuning the dwell time and distance between consecutively visited probe positions.  We add a non linear function to include a threshold effect and evaluate the accumulated damage in each part of the image as a function of scan pattern details. Together, these ingredients are able to describe qualitatively all aspects of the experimental data and provide us with a model that could guide a further optimisation towards even lower beam damage without lowering the applied electron dose. We deliberately remain vague on what is diffusing here which avoids introducing too many sample specific details. This provides hope that the model can be applied also in sample classes that were not yet studied in such great detail by adjusting higher level parameters: a sample dependent diffusion constant and damage threshold.
\end{abstract}




\end{frontmatter}



\section{Introduction}
Since the advent of the aberration-corrected scanning transmission electron microscopes (STEM) achieving atomic resolution images has become very accessible. However for many materials such as inorganic and biological materials, atomic information is very hard to gather due to electron beam damage. In the accompanying \textit{Part I} of this work \cite{a_velazco_reducing_nodate}, we showed that the beam damage in a prototypical commercial zeolite is reduced when changing the electron beam scan pattern. In particular, the alternative scan, where the probe skips over two pixels, was shown to outperform the conventional raster scan. Furthermore, a damage reduction was observed when the sample was scanned three times at a short dwell time compared to two times at a longer dwell time while adjusting the dwell time to keep the same total dose. In this part, the aim is to qualitatively model the damage observed using a minimal amount of physical parameters (diffusion constant and threshold). A model based on diffusion is chosen since the damage depends on the scan pattern indicating that there is a non-local component inducing damage depending on the time at which the neighbouring pixels are visited. However by only using the diffusion as mediator of damage, no total decrease in damage can be expected as the total applied dose is all that matters after all time effects have decayed. Therefore, a threshold is introduced below which no damage is induced. This type of behaviour has been observed by others but without combining it with diffusion \cite{johnston-peck_situ_2018,jones_managing_2018,warkentin_global_2013, cazaux_correlations_1995}. In \textit{Section \ref{SutM}} of this manuscript, we introduce the model. Next, the experimental observations from part I are used to estimate the two model parameters and to verify the applicability of the model to describe the experiments. Finally, predictions are made with respect to the damage behaviour under different scan strategies for a range of values for the diffusion constant and threshold.  These predictions could be taken as guidelines to optimise experiments for different sample classes.

%

\section{Setting up the model}\label{SutM}

\begin{figure*}
	\includegraphics[width=0.9\textwidth]{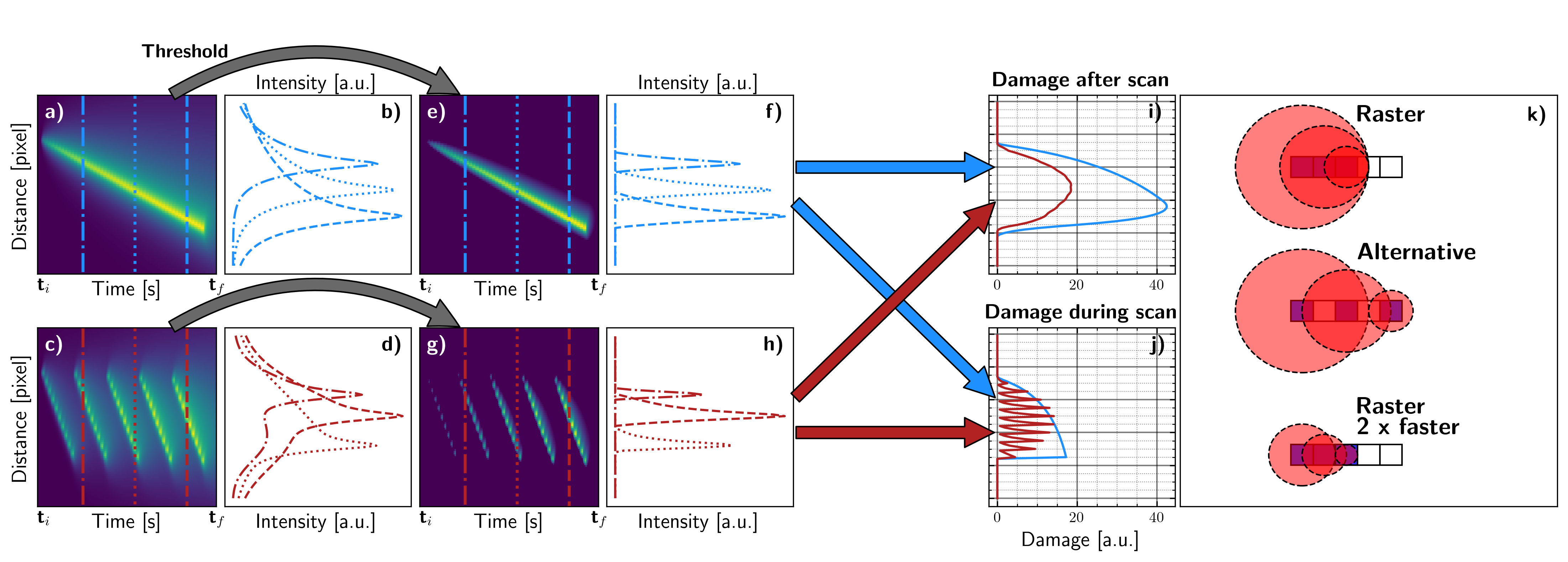}
	\caption{\label{fig_dummy} \textbf{(a)} Simplified one dimensional diffusion model for a line scan with diffusion constant 0.5 pixels/s and 50 scan points using a dwell time of ten seconds.\textbf{(b)} Diffusion profiles at 3 different times indicated on (a). \textbf{(c-d)} Same model but changing to interlaced scanning skipping five pixels. \textbf{(e,g)} Applying the threshold as in Eq. \ref{integration}. The intensity of these two images are related to the induced damage. \textbf{(f,h)} The same time frame as in (b,d) but the threshold has been applied (0.4) which significantly changes the shape of the profiles.\textbf{(i)} The damage profile of both scan patterns showing that the interlaced scan significantly reduces the simulated damage in this model. \textbf{(j)} The 'damage during scan' where the end time of the integration depends on the scan sequence, this damage is less than (i) indicating that damage gets induced after the electron beam leaves the position. For the interlaced scanning the periodicity of the scan gets imprinted on the damage profile. \textbf{(k)} Illustration of the diffusion process when raster scanning three sequential points compared to skipping pixels or scanning multiple times at a shorter dwell time}.
\end{figure*}

In an attempt to empirically describe the damage behaviour observed in the first part of this series, a model is proposed which uses a diffusion phenomenon as a mediator of beam induced damage. In this work, no assumption is made on which physical parameter is diffusing and why. Yet, these simulations and their comparison to the experiments will already reveal some of the important characteristics of the process which will help to guide further in depth research on the mechanism and how this process depends on the material and sample characteristics.
Since the thickness of electron transparent samples is small compared to their lateral dimension, we choose a 2D Ficks's diffusion model, neglecting phenomena in the 3rd dimension. 
The incoming electron beam will locally deposit energy in the sample via inelastic interaction which in turn will create a locally different state of the sample (.e.g local charging, rise in temperature, concentration of ionised species, ...). We designate the locally altered sample state by parameter $y$ and assume that diffusion of this parameter takes place (e.g. charge spreads out, heat diffuses, ionised species diffuse,...). We model the evolution of this parameter where the electron beam hits the sample as a continuous source which stays stationary at each visited probe position during the dwell time. The diffusion profile for a continuous point source in two dimension with an isotropic diffusion constant (D) is given by \cite{edward_antonian_solving_2019}:
\begin{equation}\label{CSI}
y(\bm{r},t) = \left\{
\begin{array}{lr}
0 & :t' \leq 0 \\
-\frac{I}{4 \pi D} Ei(-\frac{r^2}{4Dt'}) & : 0 < t' \leq \tau\\
\frac{I}{4 \pi D} \Bigg(Ei(-\frac{r^2}{4D(t'-\tau)})-Ei(-\frac{r^2}{4Dt'})\Bigg) & : t' > \tau
\end{array}
\right.
\end{equation}
where 
\begin{equation}
 r = \bm{r} - \bm{r_0}  \; \text{and}\; t'=t-t_0
\end{equation}
With Ei being the integral exponential function, $\tau$ the dwell time, $\bm{r_0}$ is the position of the incoming beam and $t_0$ is the time when the electron beam arrives at point $\bm{r_0}$. The derivation of Eq. \ref{CSI} can be found in \textit{supplementary materials S\ref{DCPS}}. 
The instantaneous current (I) of the parameter $y$ is related to the incoming electron beam current and the cross section for the specific inelastic excitation which depends on the material itself. We scale $I=1s^{-1}$ in the remainder of this manuscript as we are interested in a relative damage comparison between different scan conditions rather than absolute numbers. Note that $y$ at $\bm{r_0}$ goes to +$\infty$ which can be corrected for acknowledging that our electron source has a finite width where $r^2\rightarrow r^2+ \sigma^2_P$ with $\sigma_P$ the probe diameter estimated as $\sigma_P \approx$ 70 pm for atomic resolution STEM experiments. The full diffusion profile ($Y$) can be written as the superposition of the diffusion profile originating from every scanned point given by: 
\begin{equation}
Y(r,t) =  \sum_{i=1}^n y(\bm{r}, t, \bm{r_{i}}, \tau_i, t_{0,i})
\end{equation}
With $\bm{r_i}$  the scan positions visited at time $t_{0,i}$ where the electron probe stays for a time $\tau_i$ which will be referred to as the dwell time. In conventional STEM experiments this value will be the same for every scan position.   
So far, the model is linear and can not explain the observed differences in damage for different scan patterns as the total dose and dose rate remained identical. This observation calls for a nonlinear component where damage only sets in if the parameter $Y$ crosses a certain threshold. We assume the simplest hard-cut model. Physical reasons for such a threshold could be related to dielectric breakdown, phase changes, reaction rates, or many other depending on the specific meaning of the $Y$ parameter.
We now can calculate the induced sample damage $D_a(\bm{r})$ in a certain position in the sample, by applying the threshold and integrating over time from zero to a final time $t_f$:
\begin{equation}
D _{a}(\bm{r},t_f) = \int_{0}^{t_f} \max(Y(\bm{r},t)-T_h) \,dt \label{integration}
\end{equation}
where $T_h$ is the threshold. Now we hit a subtle but important point: while scanning a single image, the only damage that matters is the damage that occurred before or during the recording of a given probe position. Any damage that occurs in that position after the recording of that position will not change the intensity value in the image for that position any more and $t_f$ will be different for each scanned position. If on the other hand, a second image is recorded (and ignoring the damage that this new image will do), e.g. to evaluate the beam damage in a previous recording, then all damage occurring before is taken into account. The result will be different in general (more damage) as $t_f$ is now taken in the limit to infinity, assuming a long ($t_f >> \frac{\Delta r^2}{4D}$, with $\Delta r$ the image size) time passed between both images.
In the remainder of this manuscript we will discuss and compare the consequences of this under the labels 'damage during scan' and 'damage after scan'.
 
   
In Fig. \ref{fig_dummy}, a one dimensional toy model is represented where fifty points using a dwell time of ten seconds are scanned with two different patterns, the first being sequentially and the second being interlaced skipping four pixels on every pass until all points are visited. The diffusion constant is taken as 0.5~pixels/s with a current of  1~s$^{-1}$. The diffusion profiles for both scans at every time step are shown in Fig. \ref{fig_dummy} (a,c). There is a clear dependence of the diffusion profiles with scan pattern. The next step in the model is applying the threshold which is visualised in Fig. \ref{fig_dummy} (e,g). Depending on whether we assume 'damage during scan' or 'damage after scan' we either integrate this damage evolution at the point of leaving a certain position or at the end when all scanning effects have subdued under the threshold level. The threshold value used in this 1D model was $T_h=0.4$. By comparing the profiles in Fig. \ref{fig_dummy} (b,d) and (f,h), the thresholded intensity profiles show different shapes due to the nonlinearity in the model. From this thresholded diffusion profile, the two different damage profiles can be calculated. The 'damage after scan' and 'damage during scan' are shown in (i) and (j) respectively. For both cases, the interlaced scan induces less damage than the raster scan. Further the damage profile between the two simulations is quite different as the damage almost doubles when calculating the 'damage after scan' implying that there occurs a significant amount of damage in the point after it has been visited by the electron beam.  In this paper two different routes are investigated on how to reduce damage when keeping the total dose and the dose rate constant. An illustration on the different damage reduction methods are shown in Fig. \ref{fig_dummy}(k). The first one is shown in the middle where two pixels are skipped compared to the raster scan in an attempt to avoid damage buildup. When the probe position arrives at the end, it starts the scan with the same number of pixels skipped but the starting point is next to the first visited point. This type of scanning will be referred to as alternative scanning as explained in \textit{Part I} \cite{a_velazco_reducing_nodate}. The second method is via scanning faster over the sample with multiple passes until reaching the same total dose in the end. 

We numerically implemented eq. \ref{integration} using a time step of $\Delta t=1\;\mu s$ as a good compromise between accuracy and speed. In \textit{supplementary materials S\ref{CT}}, the convergence of the integration is investigated for the different configurations. This approach has the downside of having a rather long computation time, limiting the practical simulation on a desktop computer to maximum of $32\times 32$ scan points. This limitation will have some effect on the quantitative match between model and experiment, but the aim of this paper is to get a qualitative grasp of the essential parameters that play a role and to evaluate whether a simple model can describe the main observations in the experiment starting from as few parameters as possible (diffusion constant and threshold).
The simulation code as well as all experimental results are available on Zenodo in an attempt to make this research as reproduceable as possible and to stimulate further research \cite{d_jannis_reducing_nodate}.

\section{Estimating the diffusion constant and threshold from experiment}
\begin{figure}
	\includegraphics[width=\columnwidth]{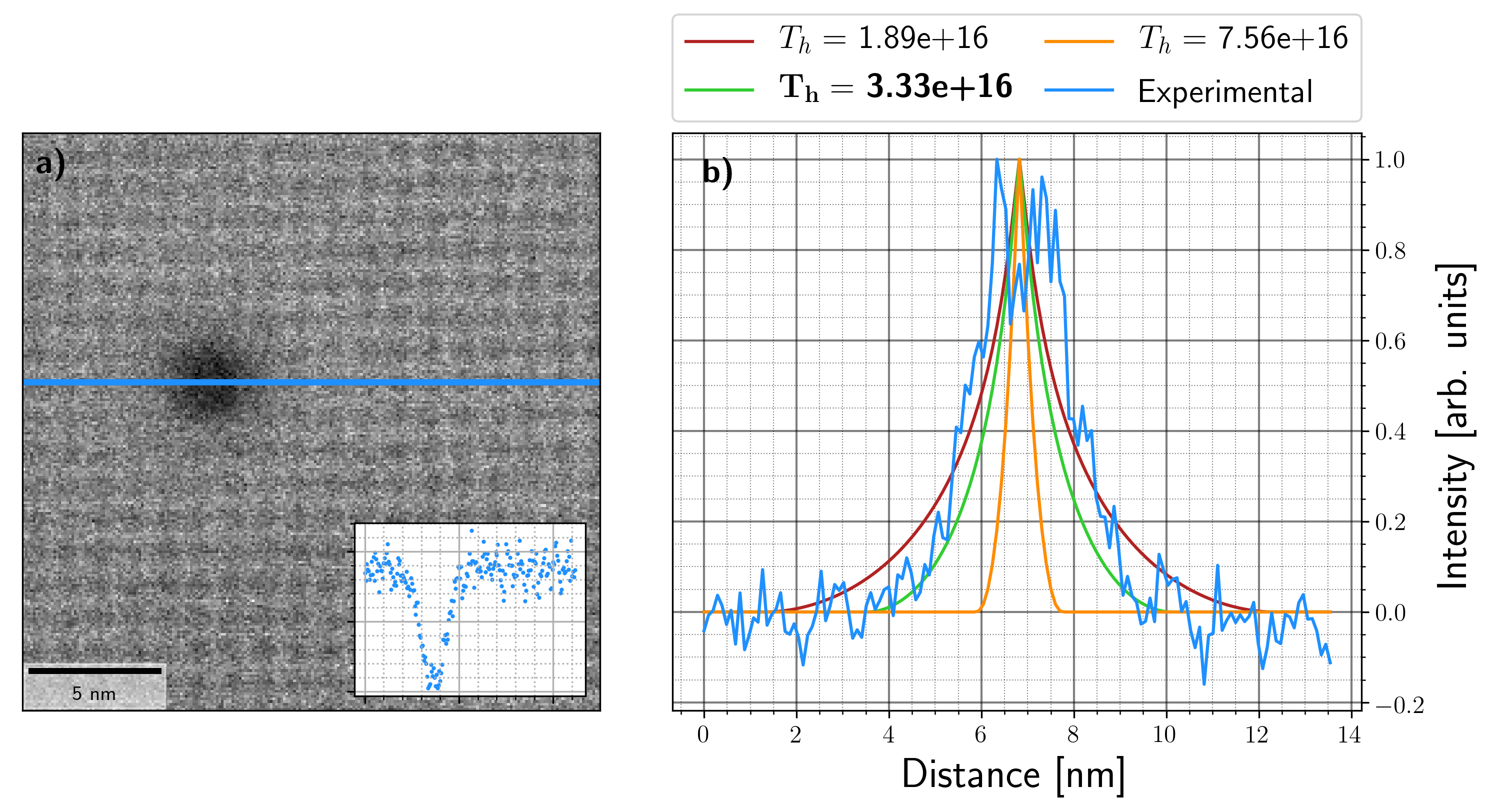}
	\caption{\label{fig_holedrilling} \textbf{(a)} Low magnification HAADF scan of the zeolite sample. A dark hole is visible which arises from the placement of a static probe on that area for 2~s. The reduced intensity indicates the damage and mass loss of the sample. \textbf{(b)} In blue, the inverse signal of (a), indicating the damage, the other colours indicate different thresholds for the simulated damage. For every simulation, the diffusion constant has been kept constant (4.5 nm$^2$/s). The simulation with threshold $3.33\times 10^{16}$ seems to best fit the experimental damage profile (green) which is indicated with the bold font.}
\end{figure}

\begin{figure*}
	\includegraphics[width=0.9\textwidth]{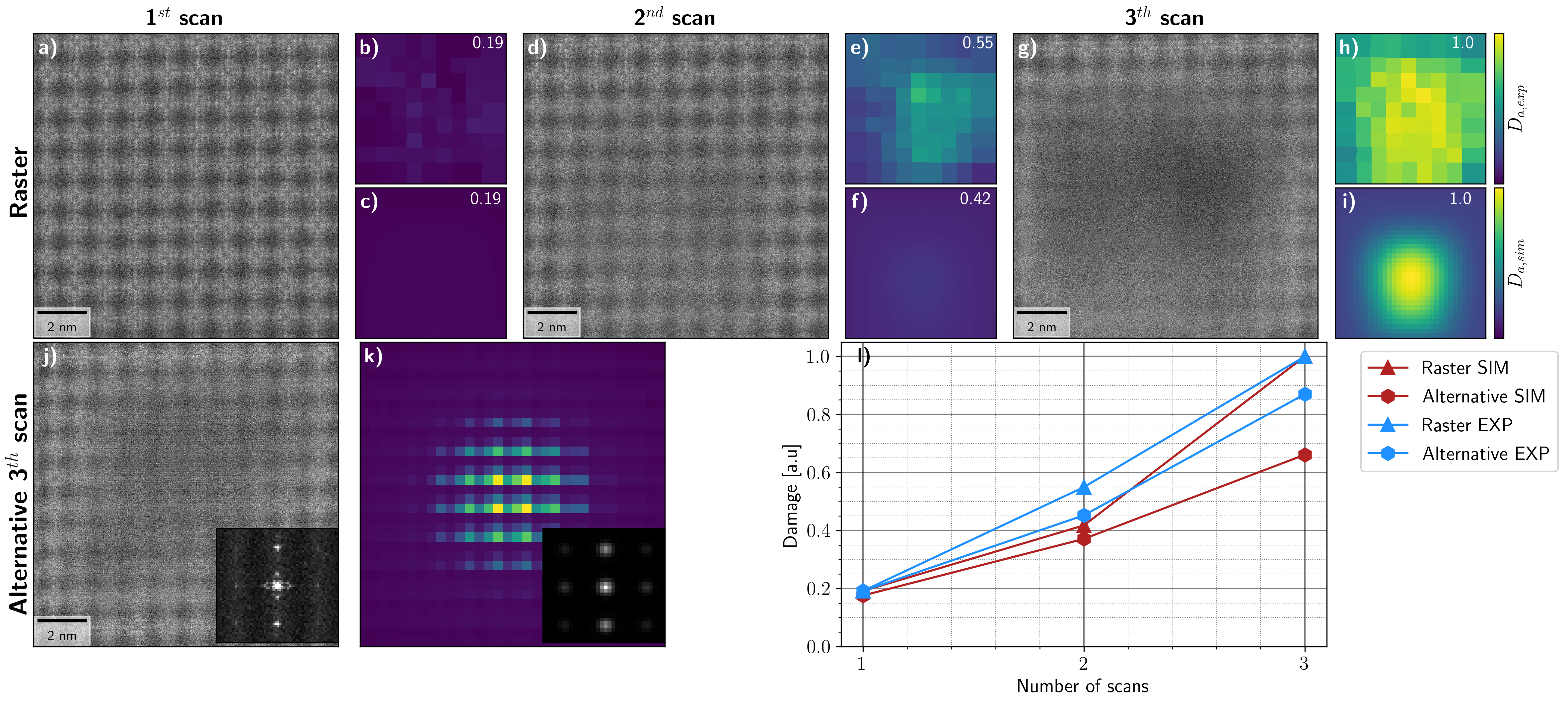}
	\caption{\label{fig_comparison_exp_sim} \textbf{(a,d,g)} 512x512 HAADF scans over the zeolite where the three scans are performed after each other which where performed with the conventional raster scanning with a dwell time of 6~$\mu$s. \textbf{(b,e,h)} The experimental damage $D_{a,exp}$ using Eq. \ref{for:ncc} which is calculated for every unit cell and interpolated over the scan area. \textbf{(c,f,i)} The simulated damage profiles using the diffusion constant of 4.5 nm$^2$/s and a threshold value of $3.33\times 10^{16}$ which was found in Fig. \ref{fig_holedrilling}.The text on the upper right indicated the total damage compared to the last scan. \textbf{(j)} The third scan of the alternative scan pattern where the induced damage is seen as a decrease of signal. On the inset the Fourier transform is seen where extra spots are seen at high frequencies. \textbf{(k)} The simulated 'damage during scan' profile for the alternative scan using the same parameters. On the inset the Fourier transform is shown where the extra spots are seen at the same position arising from the imprinting of the damage due to the alternative scanning pattern. \textbf{(l)} The simulated 'damage during scan' for the three consecutive scans (red) where the induced damage is less for the alternative compared to the raster scan which is observed in the experiment data (blue). }	
\end{figure*}

An estimate for the diffusion constant and threshold will be obtained by qualitatively fitting the model with two specific experiments:

\begin{description}
	\item [$\bullet$ ]Placing a static probe on the sample for 2 seconds.
	\item [$\bullet$ ]Scanning the sample three times with the raster and alternative method at 6~$\mu$s dwell time.
\end{description}
The experiments performed in \textit{part I} \cite{a_velazco_reducing_nodate}  are used to estimate the parameters where the current used in the two experiments was kept constant at  50~pA with an incoming electron beam of 300 kV. The sample is a commercial
Linde Type A (LTA) zeolite sample (calcium exchanged sodium aluminium silicate, Sigma Aldrich BCR-705) \cite{mayoral_atomic_2011}, which are known to be beam sensitive materials \cite{csencsits_damage_1987, treacy_electron_1987}.
The first experiment uses a static electron probe placed on the sample for 2~s. After this, a large field of view scan was performed to inspect the local damage. In  Fig. \ref{fig_holedrilling} (a), the high-angle annular dark field (HAADF) scan (1024$\times$1024 at 3~$\mu$s) is shown. Around the centre, a dark region is observed which arises from the interaction of the static probe where the lower signal indicates mass loss induced around the illuminated area. The blurring observed around the damage zone, indicates a loss of crystallinity. The first observation of the damage profile is that it is isotropic indicating the isotropy of the diffusion process despite the single crystalline nature of this region of the sample.
We estimate the diffusion constant by measuring the width of the damage profile to be $\pm$~5 nm for a probe illumination of 2~s. This leads to an order of magnitude for the diffusion constant in nm$^2$/s. In this work three different diffusion constants are shown. In the main text, the diffusion constant which has showed the best comparison with the data, $D=4.5$~nm$^2$/s  is used. In \textit{supplementary materials S\ref{ODC}}, simulations with two other diffusion constants are shown (D=45nm$^2$/s and D=0.45nm$^2$/s) in order to demonstrate the effect of this parameter on the model. The static probe experiment is used to determine the threshold for a particular diffusion constant. Hence for every diffusion constant a proper threshold could be found to describe the damage observed in the static probe experiment. In  Fig. \ref{fig_holedrilling} (b), three simulated damage profiles are shown where only the threshold is varied. From this a threshold value of $3.33\times 10^{16}$ is chosen since it is most closely resembles the experimental damage profile. 
In the remainder of the text we use:
\begin{center}
\begin{tabular}{c c}
$D=$ &  4.5 $nm^2/s$\\
$T_h=$ &  $3.33\times 10^{16}$
\end{tabular}
\end{center}

The next experiment is used to verify if the chosen diffusion constant and threshold are able to reproduce the experimental damage profiles. This is done by comparing the simulated damage profiles with the experimental data from the  three consecutive scans using the raster scan method. The type of simulation used is the 'damage during scan' method since this is the damage observed during the experiment. In Fig. \ref{fig_comparison_exp_sim}(a,d,g), the three experimental HAADF scans (see \textit{Part I} \cite{a_velazco_reducing_nodate}) are shown where more damage is seen in the third scan (g). The damage can be observed by the reduced intensity and disappearance of the crystalline structure. 
In order to quantify the total induced damage per scan the first proposition would be to track the total HAADF intensity per image. However no loss in intensity is observed indicating that the loss of mass happens locally, as seen by the darker regions in the centre of Fig. \ref{fig_comparison_exp_sim}(g). Therefore another method is used to get some quantification of damage where the damage for every unit cell is calculated by template matching it with the average undamaged structure. Hence for every unit cell the normalized cross correlation (NCC) is calculated. This method is explained in more detail in \textit{Part I} \cite{a_velazco_reducing_nodate}. In general, the HAADF intensity describes better the induced damage when no structure is visible and the total intensity decreases compared to the non-damaged area, see \textit{Section \ref{verification}} where this type of damage is observed. However when  some crystalline structure is seen, the NCC would be the preferred estimate of damage as it sets in sooner and is highly relevant when mapping the atomic structure of a material. Note that NCC provides some type of damage measure but is not an absolute measurement for damage. For instance, NCC=0 would correspond to an image that is orthogonal to the reference undamaged image which would be a unphysical situation that can not occur from damage. At the same time NCC=1 is not a good measure for 'no damage' as noise in the images will make that two images are never identical even when no damage occurs.
For this reason we scale the NCC value from its smallest (highest damage) to largest (no damage) value in a series of experiments in order to maintain some kind of qualitative and relative measure that can be compared to simulations. The following formula is chosen to be able to compare the simulation with experimental NCC values:
\begin{equation} \label{for:ncc}
D_{a,exp}=(1-D_{a,r1}) \cdot \frac{(NCC_{max}-NCC)}{NCC_{max}-NCC_{r3}}+D_{a,r1}
\end{equation}
where $D_{a,r1}$ is the damage level of the first raster scan as we cannot tell if there is already damage induced in the first scan step. Further $D_{a,exp}$ is chosen such that the damage induced in the last raster scan ($NCC_{r3}$) is equal to one which is also done for the simulated data. The information on the NCC coefficient and position of each unit cell is used to get a interpolated damage map which is shown in Fig. \ref{fig_comparison_exp_sim}(b,e,h). The text on the upper right indicates the total damage ($D_{a,exp}$) relative to the last scan. Simulations are performed with a $32 \times 32$ scan using the parameters found in the static probe experiment as shown in Fig. \ref{fig_comparison_exp_sim}(c,f,i). The text on the upper right indicates the total damage relative to the last scan. These simulations resemble the experimental data where in the first scan almost no damage is induced. After the first scan, the damage is centred as seen in the NCC maps giving an indication that the chosen parameters for diffusion constant and threshold are qualitatively representing the experimental behaviour. Another check is performed by investigating the third scan of the alternative scanning method (Fig. \ref{fig_comparison_exp_sim}(j)). From the Fourier transform (see inset of Fig. \ref{fig_comparison_exp_sim}(j)), extra spots are visible indicating a modulation of the intensity with a period of three pixels due to the alternative scan performing nine consecutive scans at 3$\times$3  different starting positions where in the first scan, the induced damage is smaller than for the later scans as discussed in more detail in \textit{supplementary materials S\ref{OGSfAS}}. The corresponding simulation of the scan pattern is shown in Fig. \ref{fig_comparison_exp_sim}(k), the same three pixel period modulation is seen in the simulation confirming the imprinting of damage due to the scan pattern. Finally, the total experimental and simulated damage for the three scans and the two scan sequences are compared. This is plotted in Fig. \ref{fig_comparison_exp_sim}(l) where the damages are scaled relatively to the third raster scan for both experiment and simulation. For the experimental data, as described previously, the damage in the first scan has been set equal to the simulated damage. For both experimental and simulated, the alternative scan method induces significantly less damage and the trend of damage buildup per scan is similar. The match between simulation and experiment is not quantitative which could be expected in view of the less than ideal match between NCC and simulated damage level and the reduced scan pattern used in the simulation due to computational constraints. Nevertheless the model seems to be able to give a reasonable qualitative description of the experiments.

\section{Verification of the model} \label{verification}
\begin{figure*}
	\includegraphics[width=\textwidth]{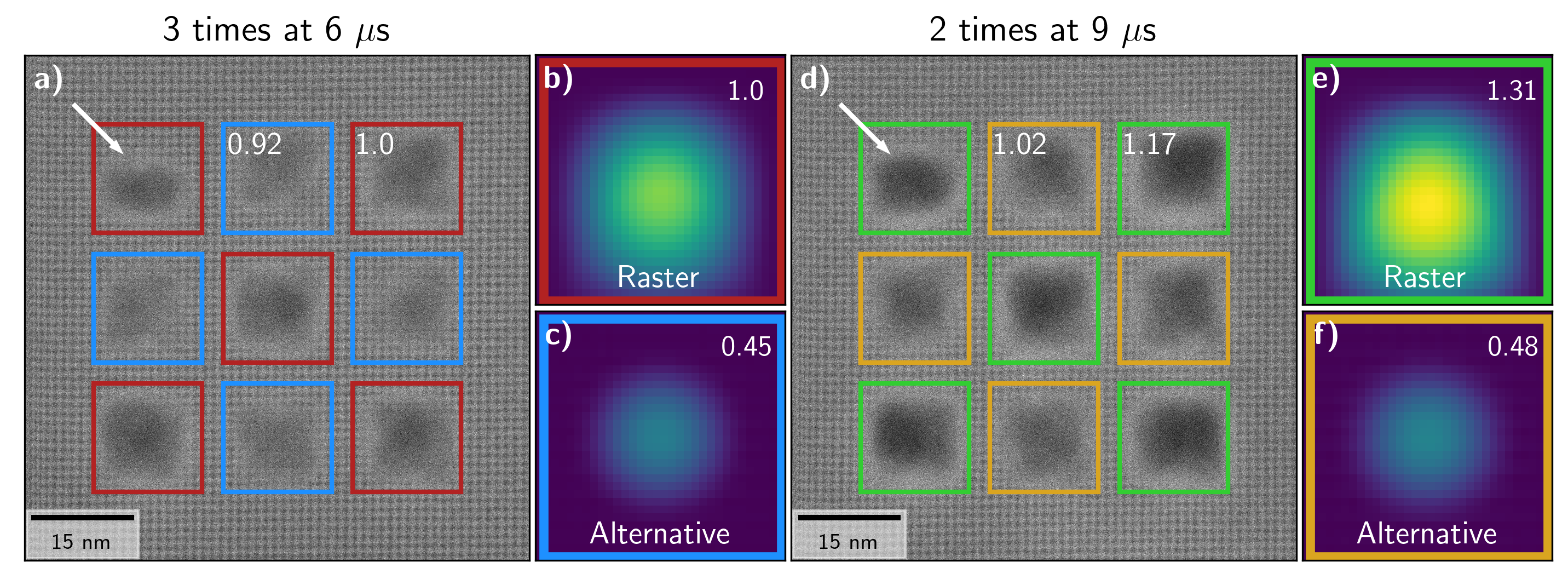}
	\caption{\label{fig_6_9} \textbf{(a)}  Large overview scan where a 3$\times$3 grid is scanned with different scanning sequences (Red=raster, blue=alternative scan), the grid was scanned three times with a dwell time of 6~$\mu$s. \textbf{(d)} The same overview scan where the grids are scanned two times with a dwell time of 9~$\mu$s (green=raster, yellow=alternative scan). In the rectangles the average HAADF intensity for each scan type relative to the raster scan at 6~$\mu$s is shown. The white arrows indicate the absence of damage which is due the unblanking of the beam at the start of the scan. \textbf{(b,c,e,f)} The simulated damage profiles using the "damage after scan " method for the four different types of scanning which are three times raster scanning (red), three times alternative scanning (blue) at 6~$\mu$s, two times raster scanning at 9~$\mu$s (green) and two times alternative scanning at 9~$\mu$s (yellow). For these the total damage relative to the 6~$\mu$s raster scan is indicated on the profiles.
	}	
\end{figure*}

In the previous section, the diffusion model and threshold are estimated from two experimental observations. These values can now be used to verify whether the model can qualitatively describe other experimental damage profiles. From the first part of this series it is clear that doing the alternative scan shows less damage than performing a raster scan when doing three consecutive scans at 6~$\mu$s. In Fig. \ref{fig_6_9} (a,d), the low magnification HAADF scans are shown where a 3x3 grid with different scanning patterns (raster and alternative) where (a) is scanned three time at 6~$\mu$s and (d) two times at 9~$\mu$s. The coloured squares indicate which grid is scanned with which pattern. In the inset of the rectangles, an estimate of the induced damage relative to the raster scan at 6~$\mu$s is shown (see \textit{Part I} for determination of average HAADF intensities)). This is done by calculating the mean HAADF intensity for the different scans. Since no crystal structure is observed, the HAADF intensity is preferred as damage indicator compared to the NCC method here. In Fig. \ref{fig_6_9} (b,c,e,f), the simulated damage patterns are shown where on the top right the relative total damage is indicated. The type of simulation in this case is the "damage after scan" since this is what we actually record in the experiment. The scan sequence is displayed at the bottom of the image. Overall, the trends are similar between the experimental and simulated data except that the alternative scan at 9~$\mu$s shows similar damage as the 6~$\mu$s raster scan in the experimental data whereas for the simulation it should be less and more similar to the 6~$\mu$s alternative scan. Further the quantitative numbers do not compare to the simulation. No clear reason for this mismatch between simulation and experiment is understood. One reason can be that the HAADF intensity underestimates the damage. Indeed the observed mass loss is also accompanied by a buildup of mass at the edges of the subimages indicating a damage process that is much further progressed which could introduce further nonlinearities in the process.
Furthermore, due to computational limitations the simulated scan pattern is restricted to $32 \times 32$ instead of the experimental $512 \times 512$ which further complicates a quantitative comparison.
\section{Alternative scan techniques}
\begin{figure} [!htb]
	\includegraphics[width=0.5\textwidth]{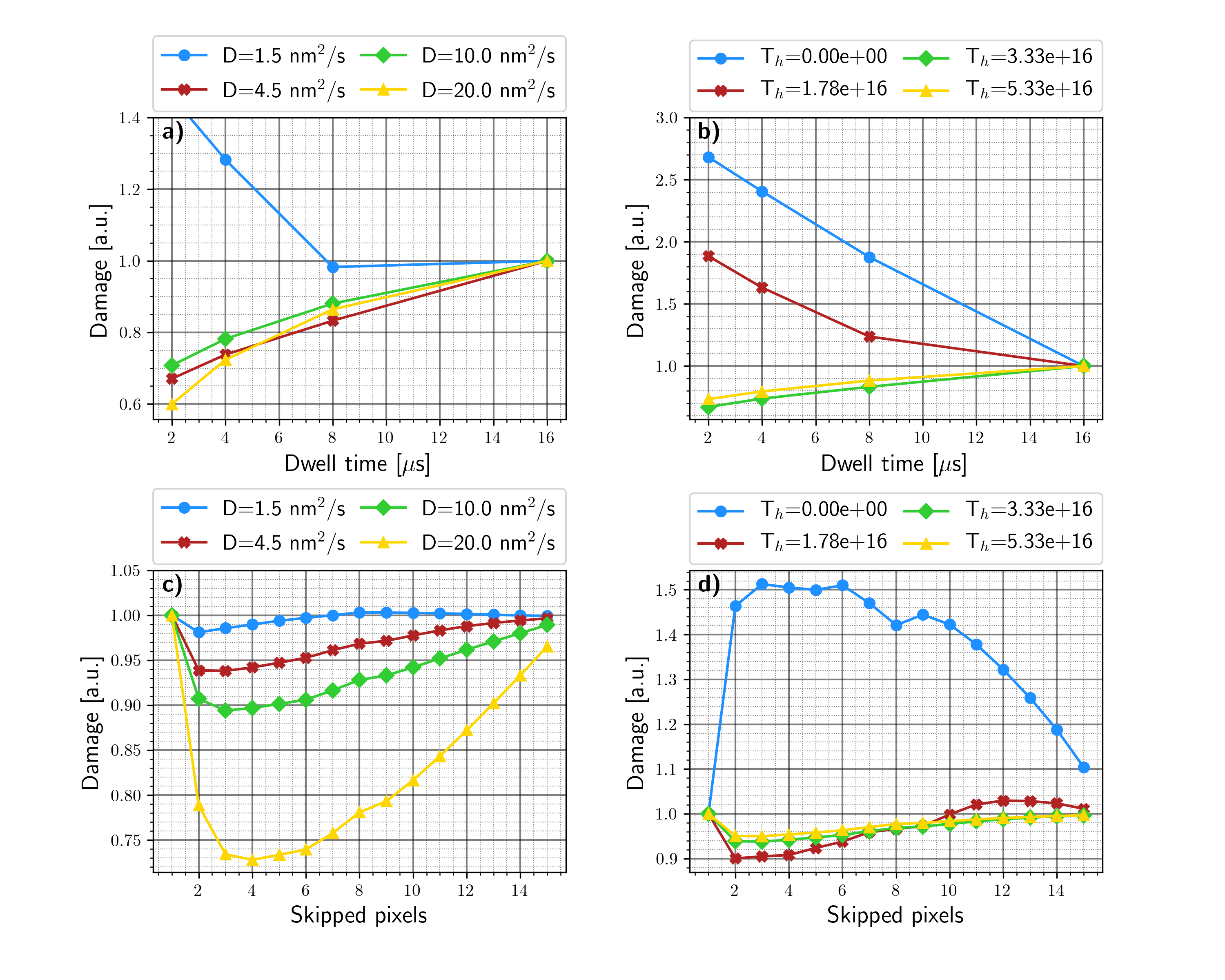}
	\caption{\label{fig_skip_mult} \textbf{(a)} The 'damage during scan' as a function of dwell time where the diffusion constant is changed and the threshold value has been kept constant ($3.33\times 10^{16}$). The pixel size is 24.2 pm and the total dose in each simulation is constant. Therefore, for the 2~$\mu$s dwell time, the total number of scans is eight where for the 16~$\mu$s one, only one scan is performed. The damage for every parameter set is normalized to the scan at 16~$\mu$s \textbf{(b)} Similar to (a) except that the threshold is varied while keeping the diffusion constant (4.5 nm$^2$/s). \textbf{(c)} The damage as a function of skipped pixels while varying the diffusion constant and keeping the threshold constant ($3.33\times 10^{16}$). \textbf{(d)} Similar to (c) but the threshold is changed while the diffusion constant is fixed. The damage is normalized for every parameter set when the number of skipped pixels is one. All these simulation show the 'damage during scan'. Note that even for zero threshold the damage is not independent on dwell time or scan pattern as we estimate the 'damage during scan' setup. For 'damage after scan' the damage will only depend on the total dose in that case as shown in \textit{supplementary materials S\ref{DASfMSaSP}}.
	} 	
\end{figure}
In this section the two methods of damage reduction, changing scan sequence and scanning multiple times, are simulated for different sets of diffusion constant and threshold. These simulation have the same pixel size as in the experiment and the scan size has been reduced to $16 \times 16$ in order to have reasonable simulation times. For the multiple acquisitions, the induced damage is calculated for four different scan configurations where the dwell time and number of scans is varied in such a manner that the dose rate and total dose stay constant. We take conventional HAADF scan dwell times which are 2, 4, 8 and 16~$\mu$s. In order to keep the total dose constant, the number of scans are respectively 8,4,2 and 1. In Fig. \ref{fig_skip_mult} (a), the 'damage during scan' as a function of dwell time is shown for different diffusion constants. In the simulation the threshold value has been kept constant at a value of $3.33\times 10^{16}$. From (a), it is seen that for every diffusion constant except for the 1.5~nm$^2$/s, the multiple scanning is advantageous compared to the single scan. In particular, the highest diffusion constant shows the largest gain in terms of damage reduction. For the 1.5~nm$^2$/s a local minimum in damage is observed when scanning two times at 8~$\mu$s. This gives insight that there are local minima and the induced damage as a function of dwell time is not monotonic indicating that scanning faster does not always reduce damage . In Fig. \ref{fig_skip_mult} (b), diffusion has been kept constant at a value of 4.5 nm$^2$/s and the threshold is varied. From this it is seen that when the threshold is low, the damage caused by multiple scanning can be worse than a single scan. This can be understood as most of the damage gets induced after the probe leaves the position and coming back to the same position enhances the damage. For these types of parameters, it is beneficial to scan only once and get the signal before the material is destroyed.  Note how this concept relates to the 'diffract-and-destroy' concept that is used in free electron laser setups for protein diffraction \cite{neutze_potential_2000, chapman_femtosecond_2011}. Regarding the reduction of beam damage, our model confirms e.g. results from L. Jones \textit{et al.} \cite{jones_managing_2018}, in this particular sample, the experiments indicate that fractionating the dose by scanning faster and increasing the number of acquisitions reduces beam damage. This effect is supported by our diffusion model for both scanning methods. 

The other beam damage reduction method used in the experiments is using a different scan sequence. Here we investigate the damage induced when varying the number of skipped pixels ranging from one to fifteen which is the maximum for a simulations of $16 \times 16$ . We choose a  dwell time of 6~$\mu$s to stay close to the experimentally relevant parameters . In Fig. \ref{fig_skip_mult} (c) the damage as a function of skipped pixels is shown where the threshold has been kept constant ($3.33\times 10^{16}$) and the diffusion constant is varied similar to Fig. \ref{fig_skip_mult} (a). For every diffusion constant there is a gain when skipping pixels which will be larger for larger diffusion constants. Further the optimum skipped pixels depends on the diffusion constant where for 20 nm$^2$/s the optimum lays at four where for 10 nm$^2$/s this is three which can be explained by the sketch of Fig. \ref{fig_dummy}  (k). There, most of the damage can be reduced by staying ahead of the diffusion front. It seems that making these steps too large gives a disadvantage since at some point the probe needs to go back to the previous point and if this happens fast, the previous diffusing intensity is still present. In Fig. \ref{fig_skip_mult} (d) the threshold is varied while keeping the diffusion constant. Here it seems that the advantage of the skipping pixels method only works when the threshold is large enough. If the threshold value is low, then there will be no advantage by skipping pixels. Note that this is the 'damage during scan' which in conventional STEM experiments is the most important. In \textit{supplementary materials S\ref{DASfMSaSP}}, the same figure is shown where the 'damage after scan' is shown. From this it is seen that the multiple scanning and skipping pixels will always be the better option to reduce this type of damage. In this section the reduction of damage was investigated by changing the scan method for different configurations of the diffusion constant and threshold. Overall, the faster scanning and skipping pixels can reduce beam damage hence one would like to minimize the dwell time and take a proper pixel skip size. However due to the finite response time of the scan coils in current instruments, the dwell time cannot be reduced below $\approx$ 1~$\mu$s as scan distortion will start to dominate the STEM images \cite{velazco_evaluation_2020}. Note that the diffusion constants and threshold are arbitrarily chosen in the vicinity of the parameters extracted from the experimental data. These parameters may not represent real materials but demonstrate how the two parameters would influence the induced damage as well as the prefered strategy to reduce beam damage. 

\section{Discussion}
The proposed model is able to reproduce the damage behaviour in a prototype commercial zeolite sample. One could wonder whether this model could also work for other types of samples by adjusting the two parameters. This needs to be further investigated in the future on other types of materials. Moreover, the estimated parameters give an indication on the type of processes that are underlying the damage. The main different types of damage mechanism are knock-on damage, thermal heating, electrostatic charging and radiolysis \cite{egerton_radiation_2004}.
For the thermal heating, the thermal diffusion constants of zeolites are reported to be in the order of $10^{11}$~nm$^2$/s which is not comparable with the estimated diffusion constant indicating that heat is not expected to be the cause for the induced damage in the current sample \cite{schnell_thermal_2013,shimonosono_thermal_2018}. For knock-on damage, atoms are displaced via the interaction of the incoming beam with the nucleus. The cross section of this type of scattering increases with the incoming electron energy where there is a threshold energy for these scattering to occur. Radiolysis introduces atomic displacements via the interaction of the incoming electrons with the electron cloud. From Ugurlu \textit{et al.} \cite{ugurlu_radiolysis_2011} it is expected that both types of interaction, knock-on damage and radiolysis, influence the damage interaction. The two types of damage mechanism are able to introduce diffusion of particles, the difficult part is to derive from these interactions its diffusion constant which is outside the scope of this work. However if both constants are known, the diffusion model can be extended to incorporate the two different processes each with its own diffusion constant making the model more accurate for the given material. This diffusion of mass is also observed in Mkhoyan \textit{et al.} \cite{mkhoyan_full_2006} where Ca ion diffused out of the illuminated area of glass $\Big($CaO-Al$_2$O$_3$)$_{0.9}$(2SiO$_2$)$_{0.1}$$\Big)$ driven by the electric field induced by the total positive charge. They estimated a diffusion constant in the order of nm$^2$/s which is similar to our results. Further in our experiments, an increase of mass is observed at the edges of the damage indicating the gain of mass in those regions which supports the idea of the migration of atoms, most likely on the surface of the sample.  This is seen in Fig. \ref{fig_6_9} (a,d) where a brighter halo is seen around the damaged areas. In order to know how this damage occurs more research is needed but the proposed model gives a tool in simulating the observed damage profiles using only two parameters which could provide insight in the detailed mechanism.
 
The model can be used to estimate the induced damage and from this, scan parameters can be chosen in order to reduce or totally remove the beam damage. This depends on the diffusion constant and threshold of the material, since in some cases the single scan is better than the multiple times scanning (see Fig. \ref{fig_skip_mult}(b)). For materials which are similar, in terms of diffusion constant and threshold, to the zeolite used in this paper, multiple scanning is better(see Fig. \ref{fig_skip_mult}(a) in red). Additionally, using other scan patterns, such as the pixel skip method, reduces damage. The two methods can be combined to even further decrease the induced damage. 
One could wonder if it would be possible to avoid damage at all by staying below the threshold during the entire scan. Although we can change the dwell time and skipping pixels for the zeolite, it seems impossible to avoid damage when keeping the total dose and dose rate equal because there is still damage induced at 2~$\mu$s close to the present scan speed limit. One could imagine that by scanning faster the damage can be eliminated, therefore a simulation using a dwell time of 200~ns is performed where the same parameters as the simulations in Fig. \ref{fig_skip_mult}(b) (green). This very fast scan gives a decrease in damage of 62\% compared to the 16~$\mu$s scan which further improves the damage reduction, however no elimination of damage is obtained via this route. It is not entirely impossible that other strategies could emerge to further evade beam damage. This requires further research and is beyond the scope of the current paper.

In the work of Nicholls \textit{et al.} \cite{nicholls_minimising_2020} a similar damage model is proposed based on diffusion. In contrast to our model, where we take the thresholded cumulative sum of the intensity, the maximum intensity which arises in the system is used. This can be seen as $D_a = \max(Y(\bm{r}))$ where at every time step $Y$ gets calculated. In \textit{supplementary materials S\ref{CCaMM}} both models are compared using the 3 times 6~$\mu$s scan. The maximum model is able to predict the improvement of the alternative scan method. However for the maximum model, most of the damage is induced at the first scan and after this only minor additional damage is induced which does not agree with the experiment. 

The diffusion constant in the proposed model would be expected to change with $k_BT$ which would lead to less diffusion at lower temperature and therefore a higher chance to cross the threshold and thus beam damage, contrary to common belief that cooling helps to avoid radiation damage especially in cryo EM. This apparent discrepancy with our model can be reconciled by assuming in this case a thermally driven degradation mechanism (e.g. melting of amorphous ice) where the diffusing parameter $y$ is the temperature itself. Any beam induced rise in temperature predicted by this model will be on top of the equilibrium temperature where indeed the chance to cross the threshold (e.g. a phase transformation temperature) can significantly increase with a higher global sample temperature. Note that the strategies of reducing beam damage here would offer an alternative over the tedious use of cryo sample holders. 



\section{Conclusion}
In this work, a qualitative model to describe beam damage in STEM experiments is proposed. The model has two parameters, a diffusion constant and a threshold and is tested on a series of experiments on a commercial Linde Type A zeolite where damage reduction was observed when changing the scan sequence. From the experimental data we were able to extract an estimate of the diffusion constant and threshold where these values can give insight into the physical process involved in the damage process. The elegance of the proposed model lies in its capacity to describe the damage process in other materials if specific values for diffusion constant and threshold can be estimated. The model can guide experiments to choose scan parameters, such as dwell time and scan sequence, in order to reduce beam damage.  
The model explains evidences scattered in literature that dose rate and dose fractionation can play a significant role in reducing beam damage, but gives a more physical handle on why this would be the case and how to update parameters in order to optimise its effect. Our simulations imply that the largest gain in damage reduction is expected from the combination of the alternative scanning with fast multiple acquisitions.
Whether the beam damage can be removed completely with a clever choice of parameters remains to be seen, but it is clear from both experimental evidence and looking at the model behaviour that significant gains in beam damage behaviour can be obtained in exchange for only a minor upgrade in the scan engine electronics.

\section*{Acknowledgements}
D.J., A.V, A.B. and J.V. acknowledge funding from
FWO project G093417N (’Compressed sensing enabling
low dose imaging in transmission electron microscopy’)
and G042920N (’Coincident event detection
for advanced spectroscopy in transmission electron microscopy’).
This project has received funding from
the European Union’s Horizon 2020 research and innovation
programme under grant agreement No 823717
ESTEEM3. The Qu-Ant-EM microscope was partly
funded by the Hercules fund from the Flemish Government.
J.V. acknowledges funding from GOA project
“Solarpaint” of the University of Antwerp.

All data discussed in this manuscript is openly available through Zenodo to stimulate further research on this topic and to improve reproducibility.

\bibliography{P2_Zotero_DJ}

\section{Supplementary Materials}
\subsection{Derivation continuous point source} \label{DCPS}
The diffusion equation is given by: 
\begin{equation}
\frac{\partial y(\bm{r},t)}{\partial t} = D \; \nabla^2 [y(\bm{r},t)]
\end{equation}
where D is the diffusion constant. The solution for an instantaneous point source is derived in \cite{pattle_diffusion_1959}
which is:
\begin{equation}
y(\bm{r},t) = \frac{Q_0}{(4 \pi D t)^{d/2}} \exp \Big(-\frac{|\bm{r}-\bm{r_0}|^2}{4Dt} \Big)
\end{equation}
where d is the number of dimensions and Q$_0$ is the quantity in our case number of electrons. The next step is to extend the model to a continuous source 
of electrons for a time t${_i}$ to t${_f}$ where $t_f-t_i = \tau$ which is the dwell time when performing STEM experiments. Lets put $t_i=0$ since it is of no use to try and describe the diffusion profile at a time before any electrons have interacted with the sample. The two dimensional diffusion profile for an incoming electron beam at constant rate I is given by:
\begin{equation}
y(\bm{r},t) = \frac{I}{4 \pi D} \int_{0}^{\min(t,\tau)} \exp \Bigg(-\frac{r^2}{4D(t-t')} \Bigg) \frac{dt'}{t-t'}
\end{equation}
where 
\begin{equation}
r = \bm{r} - \bm{r_0}
\end{equation}
The integral can be separated  into the two time regions, one where the diffusion intensity is calculated before the end of the electron beam and the other one gives the diffusion intensity after the probe is not illuminating the sample any more.
\begin{equation}\label{int_two}
y(\bm{r},t) = \left\{
\begin{array}{lr}
\frac{I}{4 \pi D} \int_{0}^{t} \exp \Bigg(-\frac{r^2}{4D(t-t')} \Bigg) \frac{dt'}{t-t'}  & : t \leq \tau\\
\frac{I}{4 \pi D} \int_{0}^{\tau} \exp \Bigg(-\frac{r^2}{4D(t-t')} \Bigg) \frac{dt'}{t-t'} & : t > \tau
\end{array}
\right.
\end{equation}

The solution of these integral can be described by using the exponential integral function.
\begin{equation}\label{Ei_f}
\int \exp \Bigg(-\frac{a^2}{b-x} \Bigg) \frac{dx}{b-x} = Ei(-\frac{a}{b-x}) + C
\end{equation}
where 
\begin{equation}
a^2 = \frac{r^2}{4D}, \: b=t
\end{equation}
By applying Eq. \ref{Ei_f} into Eq. \ref{int_two} the final result is obtained.
\begin{equation}
y(\bm{r},t) = \left\{
\begin{array}{lr}
\frac{I}{4 \pi D} \Bigg(Ei(-\infty)-Ei(-\frac{r^2}{4Dt})\Bigg) & : t \leq \tau\\
\frac{I}{4 \pi D} \Bigg(Ei(-\frac{r^2}{4D(t-\tau)})-Ei(-\frac{r^2}{4Dt})\Bigg) & : t > \tau
\end{array}
\right.
\end{equation}
The derivation of this formula is based from \cite{edward_antonian_solving_2019}.

\subsection{Convergence test} \label{CT}
\begin{figure*}[!htb]
	\includegraphics[width=\textwidth]{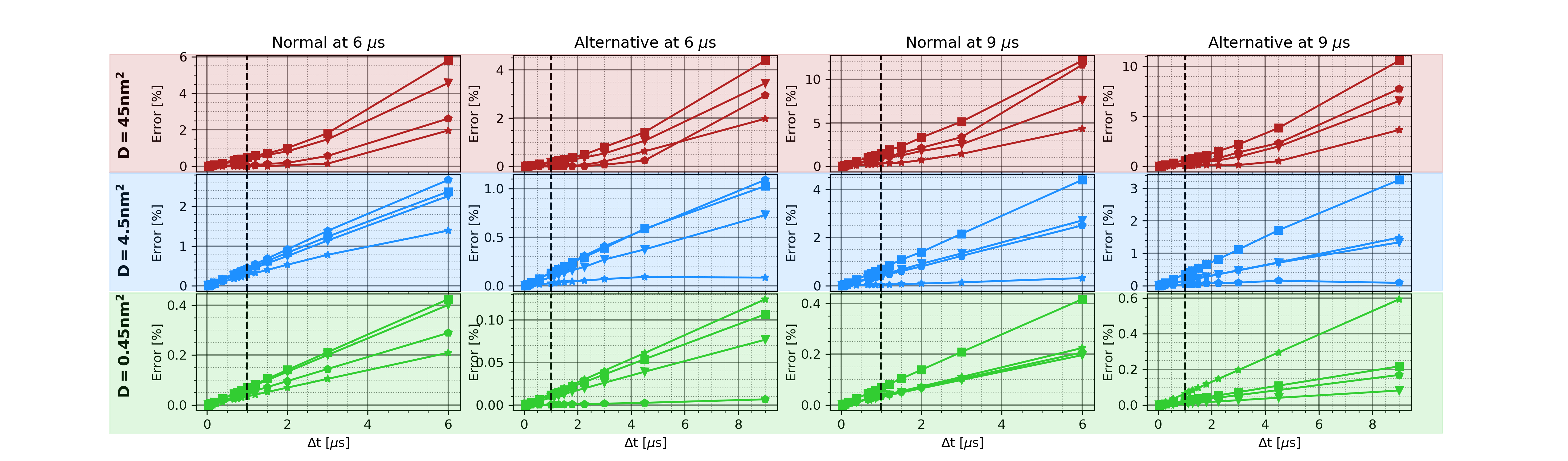}
	\caption{\label{conv_test} Every plot shows the error as a function of time steps where for large time steps, the obtained value from the integration has a larger error. For every plot different parameters such as dwell time, scan sequence and diffusion constant are varied. The main influence on the error seems to be the diffusion constant where more sampling is needed to get a small error of 2\%.  
	}	
\end{figure*}

The calculation of the total damage per scan position is performed numerical because there is no analytical solution for Eq. \ref{integration} from the main text. The result will depend on the sampling in time hence a proper time step should be taken such that the calculated damage gives a proper value and that the simulation time stays within reasonable limits. To test this, different configurations of simulations where changing the diffusion constant, scan pattern and dwell time were performed. Three different diffusion constants were taken which where used in the main part of the paper, two different dwell times are used 6 and 9~$\mu$s. In Fig. \ref{conv_test} the absolute error of the damage is shown as a function of time step for every configuration. The higher diffusion constant shows the largest relative error. From these figures we chose the size of the time step for every simulation performed to be 1~$\mu$s to have a small error (\textless 2\%) which is sufficient for the conclusions drawn from the simulations. From this it is also apparent that the necessary time step depends mainly on the diffusion constant and not so much on the dwell time and scan pattern.  

\subsection{Origin extra spots for alternative scan} \label{OGSfAS}
\begin{figure*}[!htb]
	\includegraphics[width=\textwidth]{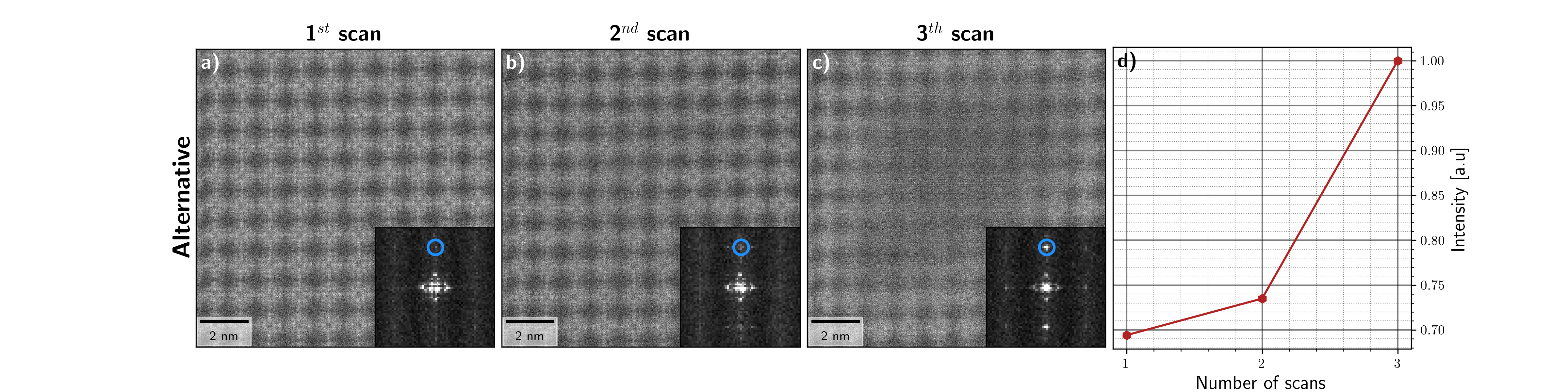}
	\caption{\label{fig_fft_alter} \textbf{(a-c)} The alternative HAADF scans over the zeolite. In the inset the Fourier transform is shown where the extra spots are visible. \textbf{(d)} The intensity of the extra spot indicated on the Fourier transforms as a function of number of scans.
	}	
\end{figure*}

In \textit{Part I} \cite{a_velazco_reducing_nodate}, the presence of the extra spots at a frequency of three pixels arise due to the sample drift. However the damage itself can imprint this type of modulation in the high-angle annular dark field (HAADF) intensity as seen in Fig. \ref{fig_comparison_exp_sim} (k) in the main text. In order to confirm that this effect is due to the damage, the three consecutive HAADF scans using the alternative scan pattern are shown in Fig. \ref{fig_fft_alter} (a-c). In the inset the Fourier transforms are shown where for the images the extra spots are visible. In Fig. \ref{fig_fft_alter} (d), the intensities of one extra spot is shown as a function of number of scans and a increase in signal occurs. If the drift during these scans is similar which is a valid statement since the drift during the acquisition was small, then this modulation is explained by the imprinting of the damage onto the HAADF signal.

\subsection{Other diffusion constants} \label{ODC}
\begin{figure*}[!htb]
	\includegraphics[width=\textwidth]{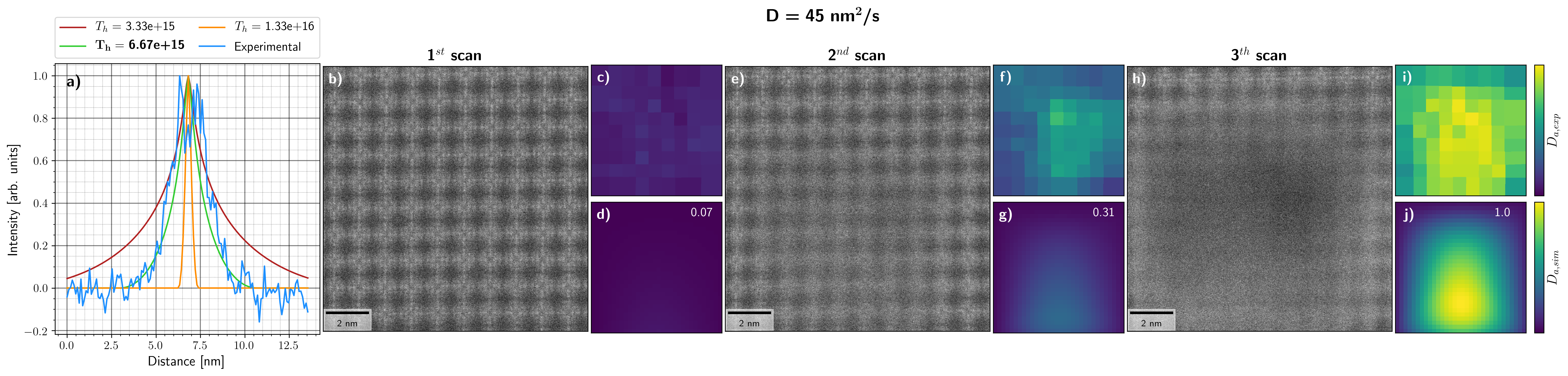}
	\caption{\label{fig_comparison_exp_sim-18} \textbf{(a)} The static probe experimental damage profile compared with simulated damage profiles while changing the threshold. The diffusion constant used for all the simulation is 45~nm$^2$/s. The threshold chosen for the raster simulation is indicated with the bold font. \textbf{(b,e,h)} 512x512 HAADF scans over the zeolite where the three scans are performed after each other which where performed with the conventional raster scanning with a dwell time of 6~$\mu$s. \textbf{(b,e,h)} The experimental damage $D_{a,exp}$ using Eq. \ref{for:ncc} (main text) which is calculated for every unit cell and interpolated over the scan area. \textbf{(c,f,i)}. The text on the upper right indicated the total damage compared to the last scan.
	}	
\end{figure*}

\begin{figure*}[!htb]
	\includegraphics[width=\textwidth]{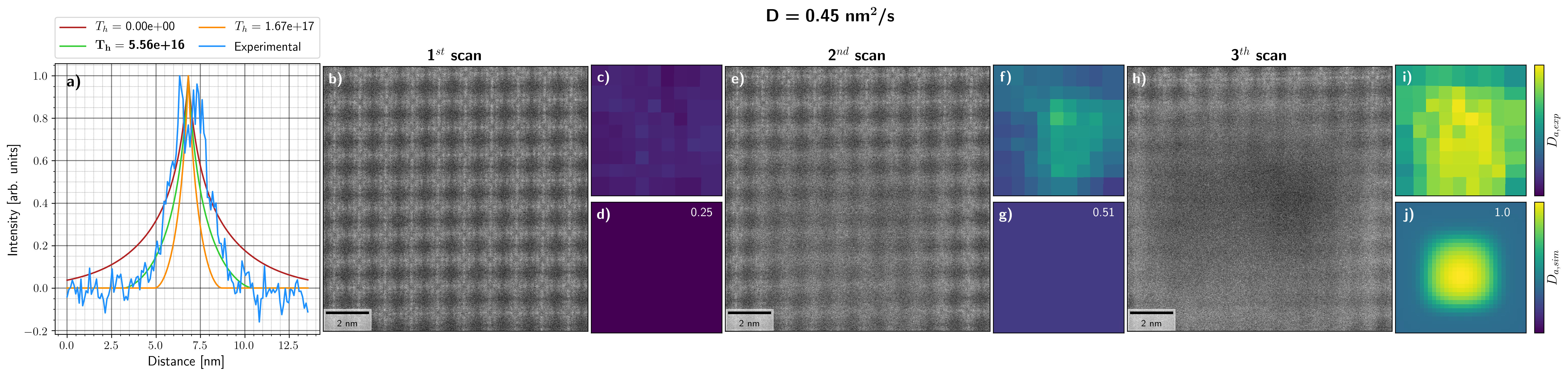}
	\caption{\label{fig_comparison_exp_sim-20} \textbf{(a)} The static probe experimental damage profile compared with simulated damage profiles while changing the threshold. The diffusion constant used for all the simulation is 0.45~nm$^2$/s. The threshold chosen for the raster simulation is indicated with the bold font. \textbf{(b,e,h)} 512x512 HAADF scans over the zeolite where the three scans are performed after each other which where performed with the conventional raster scanning with a dwell time of 6~$\mu$s. \textbf{(b,e,h)} The experimental damage $D_{a,exp}$ using Eq. \ref{for:ncc} (main text) which is calculated for every unit cell and interpolated over the scan area. \textbf{(c,f,i)}. The text on the upper right indicated the total damage compared to the last scan.
	}	
\end{figure*}

In this section, two other diffusion constants are used to perform the static probe and raster scan simulations. The two diffusion constants used are 45 nm$^2$/s and 0.45 nm$^2$/s since they are one order smaller and larger than the diffusion constant used in the main text. For the estimation of the threshold, the same procedure is used where the damage profile of the static probe is matched with the experimental data. In Fig. \ref{fig_comparison_exp_sim-18} (a) and Fig. \ref{fig_comparison_exp_sim-20} (a) show respectively the experimental and simulated damage profiles for the 45 and 0.45 nm$^2$/s. The threshold which matches the best is indicated with the bold font. For these two parameters, the simulation on the raster scan is performed. In Fig. \ref{fig_comparison_exp_sim-18}, the main difference is seen in the damage profile of the third scan where in the simulated profile, the damage is mainly seen at the bottom which is not observed in the experiment. For Fig. \ref{fig_comparison_exp_sim-20}, the third damage profile is centred but the second scan induces a constant damage on the simulated pattern (g) where in the experimental the damage already shows some preference to occur in the centre. Since the diffusion constant of 4.5 nm$^2$/s shows a better comparison of the observed damage, this one is used in the main text.

\subsection{Damage after scan for multiple scanning and skipping pixels} \label{DASfMSaSP}
\begin{figure*}[!htb]
	\includegraphics[width=\textwidth]{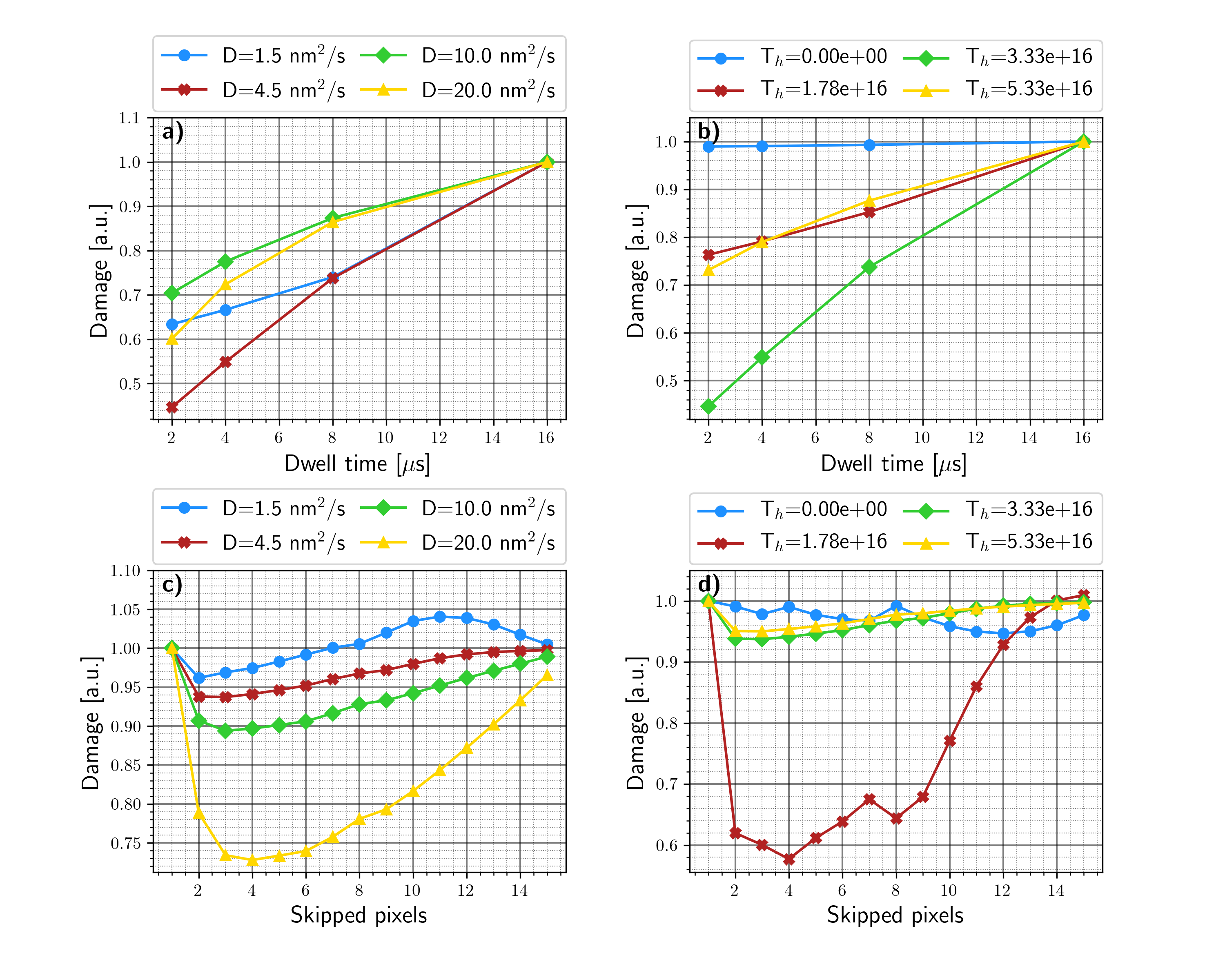}
	\caption{\label{fig_damage_after_scan}  \textbf{(a)} The 'damage after scan' as a function of dwell time where the diffusion constant is changed and the threshold value has been kept constant ($3.33\times 10^{16}$). The pixel size is 24.2 pm and the total dose in each simulation is constant. Therefore, for the 2~$\mu$s dwell time, the total number of scans is eight where for the 16~$\mu$s one, only one scan is performed. \textbf{(b)} Similar to (a) except that the threshold is varied while keeping the diffusion constant (4.5~nm$^2$/s). \textbf{(c)} The damage as a function of skipped pixels while varying the diffusion constant and keeping the threshold constant ($3.33\times 10^{16}$). \textbf{(d)} Similar to (c) but the threshold is changed while the diffusion constant is fixed. All these simulation show the 'damage during scan'.
	}	
\end{figure*}

Fig. \ref{fig_damage_after_scan} shows similar data as in Fig. \ref{fig_skip_mult} in the main text. However the difference in this figure is the method used for calculating the damage. In Fig. \ref{fig_skip_mult}, the damage is calculated during the scan whereas in Fig. \ref{fig_damage_after_scan}, the damage is calculated after the scan. From Fig. \ref{fig_damage_after_scan}(a-b), it is seen that the multiple dwell time always is the better option when wanting to reduce the total damage induced during a scan. In Fig. \ref{fig_damage_after_scan}(b) the zero threshold simulation shows that when no threshold would be introduced in the model, that the multiple scanning technique would not introduce less total damage in the sample as expected. When checking the multiple pixel skipping in Fig. \ref{fig_damage_after_scan}(c-d), the advantage for skipping pixels to reduce the total induced damage is also clear.

\subsection{Comparison cumulative and maximum model} \label{CCaMM}
\begin{figure*}[!htb]
	\includegraphics[width=\textwidth]{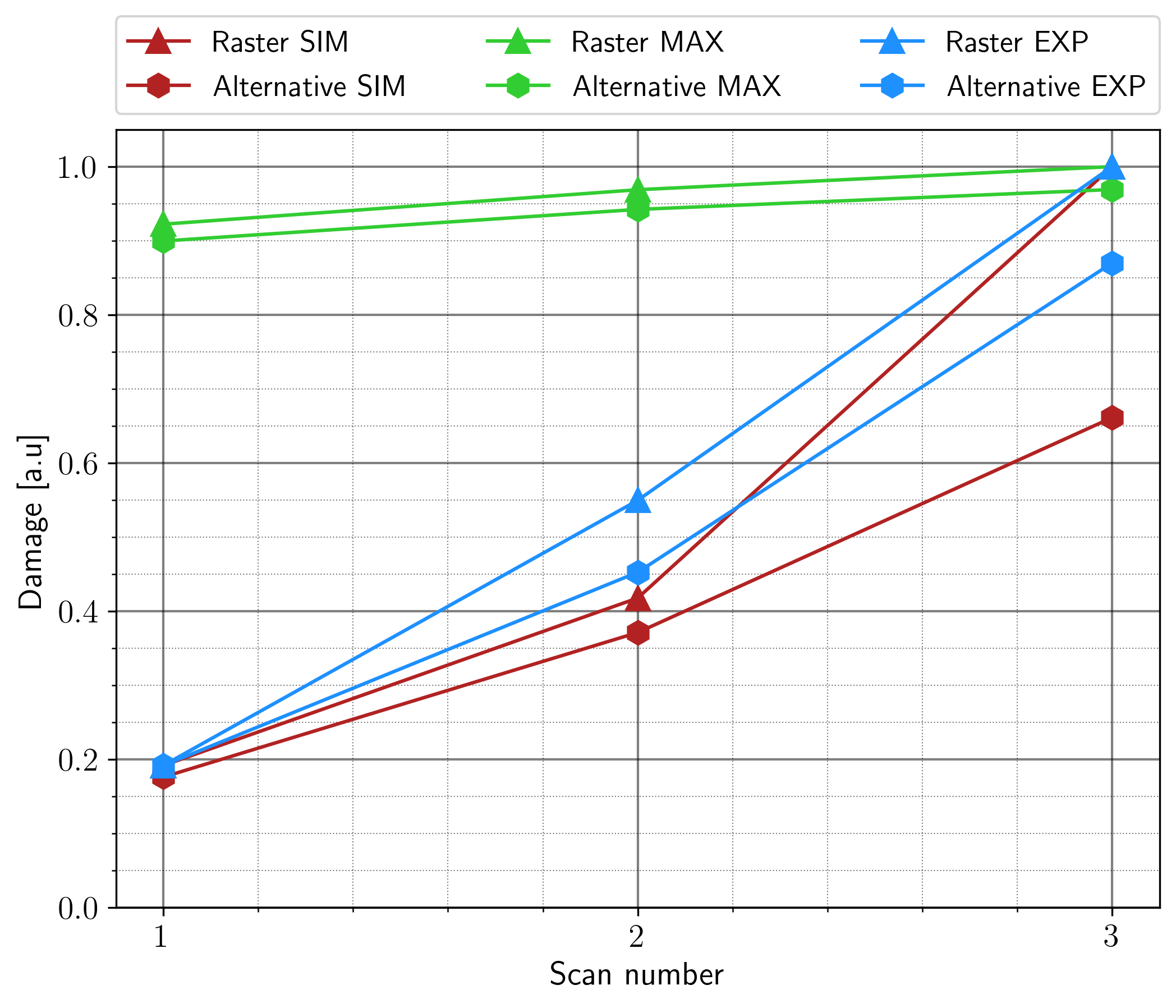}
	\caption{\label{fig_comparison_max_cum} Similar plot as in Fig. \ref{fig_comparison_exp_sim}(l) where in blue, the experimental damage per scan for the 6~$\mu$s dwell time is shown. The damage is calculated with the NCC template matching method. In red the simulated damage profiles using our model. In green, the damage when the maximum value is taken into account.
	}	
\end{figure*}

In this section, the model as described in Nicholls \textit{et  al.} \cite{nicholls_minimising_2020} is implemented by taking the maximum value of the intensity (or beam influence). The comparison with our model and the experimental determined values is shown in Fig \ref{fig_comparison_max_cum} where in green the results from the maximum model is shown. From this it is seen that the using the maximum model, most of the damage is already induced in the first scan which is not observed in the experimental data. Hence our model reproduces the experimental results better than the maximum model. However the model qualitatively predicts that the alternative scan would result in less damage.

\end{document}